\documentstyle[12pt]{article}
\def\be{\begin{eqnarray}}
\def\ee{\end{eqnarray}}
\def\vep{\varepsilon}
\def\gt{$\Gamma(t\to bW)$}
\begin{document}
\begin{flushright}
{\bf hep-ph/9503444}\\
{\bf TTP95-15}\footnote{The complete postscript file of this
preprint, including figures, is available via anonymous ftp at
ttpux2.physik.uni-karlsruhe.de (129.13.102.139) as
/ttp95-15/ttp95-15.ps or via www at
http://ttpux2.physik.uni-karlsruhe.de/cgi-bin/preprints/
Report-no: TTP95-15}\\
{\bf March 1995}
\end{flushright}
\vspace*{2cm}
\centerline{\bf TWO-LOOP LIGHT QUARK CORRECTIONS }
\vspace*{0.035truein}
\centerline{\bf TO THE TOP WIDTH\footnote{{\noindent
Dedicated to  Professor K.~Zalewski to honor his 60th Birthday.}}}
\vspace*{0.37truein}
\centerline{\footnotesize ANDRZEJ CZARNECKI}
\vspace*{0.015truein}
\centerline{\footnotesize\it
Institut f\"ur Theoretische Teilchenphysik,
Universit\"at Karlsruhe}
\baselineskip=10pt
\centerline{\footnotesize\it  D-76128 Karlsruhe, Germany}
\vspace*{20mm}

\begin{abstract}
A method of computing two-loop fermionic contributions to
the width of a heavy quark is described.
An analytical formula for this effect in the limit of mass of the
quark much larger than the decay products is obtained for the first
time.  The result confirms previous numerical studies.
\end{abstract}

\vspace*{10mm}
\section{Introduction}
\noindent
The recently discovered top quark attracts attention of many
physicists who regard its extraordinarily large mass as a hint of its
connection with ``new physics''.  One consequence of the unusually big
mass is the large width of the top quark $\Gamma_t$.  This quantity is
certainly sensitive to possible exotic particles with which top can
interact and will be subject of future precision experimental studies.
It is important to know the standard model prediction for $\Gamma_t$
as precisely as possible and a lot of effort has been invested in
studies of quantum  effects which modify it.  In the framework of
the standard model the transition of the top quark into a bottom quark
and a $W$ boson is by far the dominant decay mode and it has been the
focus of the recent studies.  In particular, one-loop QCD corrections
have been found to reduce $\Gamma(t\to bW)$ by about
10\% \cite{jk2}. Recently the first step has been made towards numerical
calculation of two-loop strong corrections to this reaction, namely
the subset of diagrams containing a fermion loop or an additional fermion
pair in the final state \cite{smith}. The relevant diagrams are shown
in fig.~1.

\begin{minipage}{13.cm}
\[
\hspace*{4mm}
\mbox{
}
\]
\vspace*{1.5cm}
\hspace*{55mm}\raisebox{-10mm}{(a)}
\vspace*{-10mm}
\[
\mbox{
\hspace*{-5mm}
\begin{tabular}{cc}
&\hspace*{10mm}
\end{tabular}
      }
\]
\hspace*{30mm}\raisebox{-10mm}{(b)} \hspace{40mm}\raisebox{-10mm}{(c)}

\vspace*{5mm}
\hspace*{-2mm}
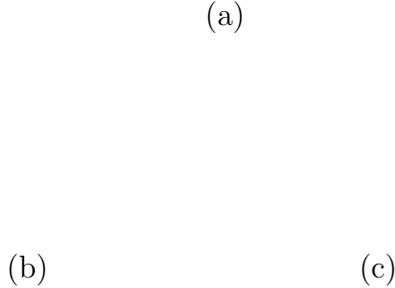
{\small Figure 1: Diagrams with virtual and real fermions contributing to
top's width.}
\vspace*{12mm}
\end{minipage}

Calculation of the fermionic contributions is particularly
important because it helps to establish the correct mass scale at
which the strong coupling constant should be taken for the calculation
of one-loop corrections.

We note that the relevant expansion parameter for the
calculations of the decay width  $\Gamma(t\to bW)$ is $m_W^2/m_t^2
\hspace*{.5mm}\raisebox{-1.2mm}{\footnotesize $\sim$}
\hspace*{-2.4mm}\raisebox{.3mm}{$<$}0.2$.  Therefore taking $m_W=0$,
as well as neglecting masses of all fermions except the decaying top
quark
as in ref.~2 is
a very good approximation.
The purpose of this paper is to show a method
in which the fermionic subset of two-loop corrections can be calculated
analytically.

\section{Real and virtual contributions}
\noindent
The tree level decay rate of a top quark into massless $W$ and $b$ is
(we take the relevant Kobayashi-Maskawa matrix element equal 1)
\be
\Gamma^0={G_Fm_t^3\over \sqrt{2}}(2\pi)^{2-D}R_2
\ee
where $m_t$ is the pole mass of the top quark (we shall use it as the
unit of mass throughout this calculation)
and $R_2$ denotes the volume of the two-body phase space
\be
R_2\equiv {\pi^{1-\vep}\Gamma(1-\vep) \over 2\Gamma(2-2\vep)}
\ee
In order to regularize infrared and ultraviolet divergences we perform
the calculations in $D=4-2\vep$ dimensions.  In particular we need to
know the one-loop gluonic corrections to \gt\   including terms ${\cal
O}(\vep)$. We find (neglecting terms containing the Euler constant
$\gamma_E$ and $\ln 4\pi$ which vanish in the final result)
\be
\Gamma^1=-{G_FR_2g_s^2C_F \over \sqrt{2}2^{2D-3}\pi^{{3\over 2}D-2}}
\left[
-{5\over 2}+{2\over 3}\pi^2
+\left(-{61\over 4}+{5\over 3}\pi^2+8\zeta(3)\right)\vep
\right]+{\cal O}(\vep^2)
\ee
where $C_F=4/3$ is the $SU(3)$ color factor.
We can now proceed to the calculation
of contributions of a single flavor of light quarks
to \gt.  We first consider the effect of the virtual correction to the
vertex, as shown in fig.~1a.  The contribution of a massless quark
loop to the gluon propagator is (see e.g.~\cite{muta})
\be
\Pi^{\mu\nu}&=&\Pi(q^2)\left(q^2 g^{\mu \nu} - q^\mu q^\nu \right)
\nonumber\\
\Pi(q^2)&=& - {ig_s^2  \over (4\pi)^{D/2}}
{2-2\vep \over 3-2\vep}  \Gamma(\vep)B(1-\vep, 1-\vep)
(-q^2)^{-\vep}
\label{eq:vacpol}
\ee
The vertex diagram can now be calculated exactly in any dimension
\be
\Gamma^V&=& -{G_F C_F g_s^4  R_2\over \sqrt{2}2^{3D-4} \pi^{2D-2} }
{(1-\vep)(2-9\vep+23\vep^2-30\vep^3+16\vep^4)\over
3(1-2\vep)^2 (1-3\vep)(2-3\vep)(3-2\vep)}
\nonumber\\
&&\times{\Gamma^2(-\vep)\Gamma(\vep)\Gamma(2\vep)\Gamma(-4\vep)\over
\Gamma(-2\vep)\Gamma(-3\vep)}
\nonumber\\
\lefteqn{\approx {G_FR_2\over \sqrt{2}}
{g_s^4C_F\over 2^{3D-4}\pi^{2D-2}}
        }
\nonumber\\ &&
\times\left[
{1\over12\vep^3}+{11\over36\vep^2}+
\left({157\over108}+{5\pi^2\over72}
\right){1\over \vep}
+{3733\over648} +{55\over216}\pi^2 +{11\over18}\zeta(3)
\right]
\ee
Similarly,  the correction to the external quark leg gives
\be
\Gamma^{Z}&=&
-{G_FR_2\over \sqrt{2}}
{g_s^4C_F\over 2^{3D-5}\pi^{2D-2}}
{\vep(1-\vep^2)\over 3(1-3\vep)(2-3\vep)}
{\Gamma(\vep)\Gamma^2(-\vep)\Gamma(2\vep)\Gamma(-4\vep)\over
\Gamma(-2\vep)\Gamma(-3\vep)}
\nonumber\\
&\approx &{G_FR_2\over \sqrt{2}}
{g_s^4C_F\over 2^{3D-4}\pi^{2D-2}}
\left(
 {1\over 4\vep^2} + {9\over 8\vep} + {59\over 16} +{5\pi^2\over24}
\right)
\ee

It is more difficult to calculate the effect of real quarks in the
final state shown in fig.~1b,c.  Especially the square of the
amplitude of the gluon emission off the decaying quark and
the interference
between amplitudes with emissions from both quarks are troublesome.
This is because the integration over
the four particle phase space in the presence of the massive
propagator in the diagrams leads in $D$ dimensions to hypergeometric
functions.  However, for our aims it is sufficient to expand the
result in a Laurent series in $\vep$.  We find
\be
\lefteqn{\Gamma^R=-{G_FR_2\over \sqrt{2}}
{g_s^4C_F\over 2^{3D-4}\pi^{2D-2}}
        }
\nonumber\\ &&
\times    \left[
      {1\over12\vep^3} + {5\over9\vep^2}
       +\left( {737\over216} - {11\over72}\pi^2\right){1\over \vep}
       + {19985\over1296} - {14\over27}\pi^2 - {73\over18}\zeta(3)
    \right]   +{\cal O}(\vep )
\nonumber\\
\ee

\section{Results and summary}
\noindent
It is interesting to look closer at the cancellation of divergences among
the diagrams calculated in the previous section.  The most singular,
$1/\vep^3$ terms, which have a purely infrared origin,
vanish after summing the vertex and the real radiation diagrams (in
fact only the interference of the two real radiation amplitudes is
relevant at this level).  The sum of the two formulas
$\Gamma^V+\Gamma^R$ contains
$1/\vep^2$ poles which are cancelled after adding the external leg
correction $\Gamma^Z$.
The singularity of the resulting formula is proportional
to the finite part of the one-loop gluonic correction $\Gamma^1$
and is removed
by expressing the one-loop result in terms of the unrenormalized
coupling constant.  Of course at this stage we have the freedom of
choosing the finite normalization of $\alpha_s$.  Various choices have
been discussed recently \cite{smith,su94}.   In order to better
understand the terminology involved we look again at the gluon vacuum
polarization in eq.~(\ref{eq:vacpol}).  Expanding the numerical factor
in this formula in $\vep$ we get
\be
\Pi(q^2)=- {4ig_s^2  \over (4\pi)^{D/2}}
\left({1\over 6\vep}+{5\over 18}+{\cal O}(\vep)\right)
(-q^2)^{-\vep}
\label{eq:ren}
\ee
The $\overline{\rm MS}$ scheme corresponds to taking only the
divergent part of this expansion for the renormalization of the
coupling constant.  In this scheme our result reads
\be
\Gamma^{\rm ferm}(t\to Wb)= \left({\alpha_s\over \pi}\right)^2
{N_f\over 3}\Gamma^0
\left[
 -{ 8\over 9} + {23\over 9}\zeta(2) + 2\zeta(3) \right]
\ee
where $\zeta(2)=\pi^2/6$ and $\zeta(3)\approx 1.2020569$, and $N_f$ is
the number of light flavors of quarks.
The authors of ref.~2 recommend using the so-called $V$ scheme \cite{BLM}
which in our case amounts to taking $-q^2=1$ in eq.~\ref{eq:ren}
(we use $m_t$ as the unit of  mass).  Our final result then becomes
\be
\Gamma^{\rm ferm}(t\to Wb)= \left({\alpha_s\over \pi}\right)^2
 {N_f\over 3}\Gamma^0
 \left[
{1\over 2} + {\zeta(2)\over 3} + 2\zeta(3) \right]
\label{eq:main}
\ee
Evaluation of the square bracket gives 3.452425\ldots\  which
confirms the numerical result given in ref.~2
\be
2.54\left({2\pi^2\over 9}-{5\over 6}\right)=3.45\ldots
\ee

The analytical result given in eq.~(\ref{eq:main}) describes the
two-loop ${\cal O}(N_f\alpha_s^2)$ correction to the decay of the top
quark into massless $b$ quark and $W$ boson.  It is possible to extend
this calculation to include the effects of both masses.  In order to
avoid difficulties connected with integrals over a four particle
massive phase space I would recommend employing the method of
asymptotic expansions (see e.g.~\cite{smirnov}).  This would reduce
the problem to calculating derivatives of the amplitudes with respect
to $m_b$ and $m_W$ in the massless limit.  Therefore, one would obtain
the same integral structures with possible higher powers of
propagators.

\newpage
\section*{Acknowledgements}
\noindent
I gratefully acknowledge helpful discussions with M.~Buza,
M.~Je\.zabek, and Q.P.~Xu.
This research was supported by a grant BMFT 056KA93P.

\end{document}